# Ammonium-, Phosphonium- and Sulfonium-Based 2-Cyanopyrrolidines for Carbon Dioxide Fixation

Vitaly V. Chaban,[1] Nadezhda A. Andreeva,[1,2] and Iuliia V. Voroshylova[3]

(1) P.E.S., Vasilievsky Island, Saint Petersburg 190000, Russian Federation.

(2) Department of Physics, Peter the Great Saint Petersburg Polytechnic University, Saint Petersburg, Russian Federation.

(3) LAQV@REQUIMTE, Faculdade de Ciências, Universidade do Porto, Departamento de Química e Bioquímica, Rua do Campo Alegre, 4169-007 Porto, Portugal.

**Abstract**. The development of carbon dioxide ($CO_2$) scavengers is an acute problem nowadays because of the global warming problem. Many groups around the globe intensively develop new greenhouse gas scavengers. Room-temperature ionic liquids (RTILs) are seen as a proper starting point to synthesize more environmentally friendly and high-performance sorbents. Aprotic heterocyclic anions (AHA) represent excellent agents for carbon capture and storage technologies. In the present work, we investigate RTILs in which both the weakly coordinating cation and AHA bind $CO_2$. The ammonium-, phosphonium- and sulfonium-based 2-cyanopyrrolidines were investigated using the state-of-the-art method to describe the thermochemistry of the $CO_2$ fixation reactions. The infrared spectra, electronic and structural properties were simulated at the hybrid density functional level of theory to characterize the reactants and products of the chemisorption reactions. We conclude that the proposed $CO_2$ capturing mechanism is thermodynamically allowed and discuss the difference between different families of RTILs. Quite unusually, the intramolecular electrostatic attraction plays an essential role in stabilizing the zwitterionic products of the $CO_2$ chemisorption. The difference of chemisorption performance between the families of RTILs is linked to sterical hindrances and nucleophilicities of the α- and β-carbon atoms of the aprotic cations. Our results are supported by the previous experimental $CO_2$ sorption measurements and systematically extend their scope.

TOC Image

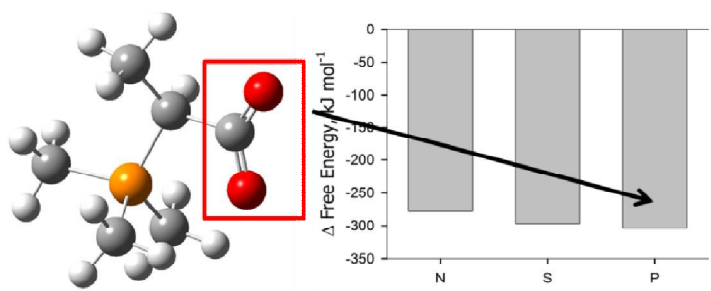

**Introduction**

Energetic and environmental issues are at the center of attention nowadays. The combustion of fossil fuels still is and will really continue to be the major source of energy, providing globally three-fourths of all our energy demands.[1] Carbon dioxide ($CO_2$) is a principal gaseous outcome of the fuel-burning process. As it is recognized by scientific and industrial communities, human-caused $CO_2$ is a major contributor to global warming. Timely actions are needed to avoid excessive $CO_2$ concentrations and, therefore, reduce long-term ecological consequences.

The research efforts must be devoted to solving two major problems. First, it is essential to implement energy-saving strategies for the efficient energy consumption. Development of the alternative renewable energy sources, such as hydroelectric, solar, tidal, wind, geothermal ones, is underway. Second, robust methods for $CO_2$ capture, storage, and utilization must be elaborated and industrially implemented. The capture of $CO_2$ using room-temperature ionic liquids (RTILs) and their task-specific derivatives is an interesting and potentially viable option.[2-7]

The most promising RTILs in the context of $CO_2$ sorption are normally composed of bulky organic cations and chemically active anions, such as aprotic heterocyclic anions (AHA). Their bulkiness and asymmetry sterically prevent RTILs from crystallizing at the ambient temperature.[8] Higher conformation flexibility of the particles in the liquid aggregate state favors a more robust performance of the substance as a gas greenhouse gas scavenger. The inability of the cation and the anion to attain a distinct electrostatically-driven coordination pattern makes both of them prospective chemical sorbents for $CO_2$. The employment of the weakly coordinating ions is a key prerequisite for the observation of the competitive gas sorption values.

The presently applied RTILs are represented by very different chemical structures, ranging from heterocycles to quaternary ammonium- and phosphonium-based, ternary sulfonium-based families. RTILs, as a rule, possess fine-tunable physical and chemical properties. The latter makes them attractive for a number of applications, in which traditional solvents perform poorly.

The most relevant physical-chemical properties of RTILs for $CO_2$ capture are vanishingly low vapor pressure,[9,10] high thermal and chemical stabilities.[11,12] However, a smaller $CO_2$ sorption capacity (as compared to industrially implemented alkanolamines)[13,14] hinders large-scale applications of these compounds. This drawback can arguably be overcome by using the chemisorption potential of the functionalized task-specific RTILs and the investigation of the chemisorption potential of the already existing structures, such as AHA, quaternary and ternary cations (Figure 1).

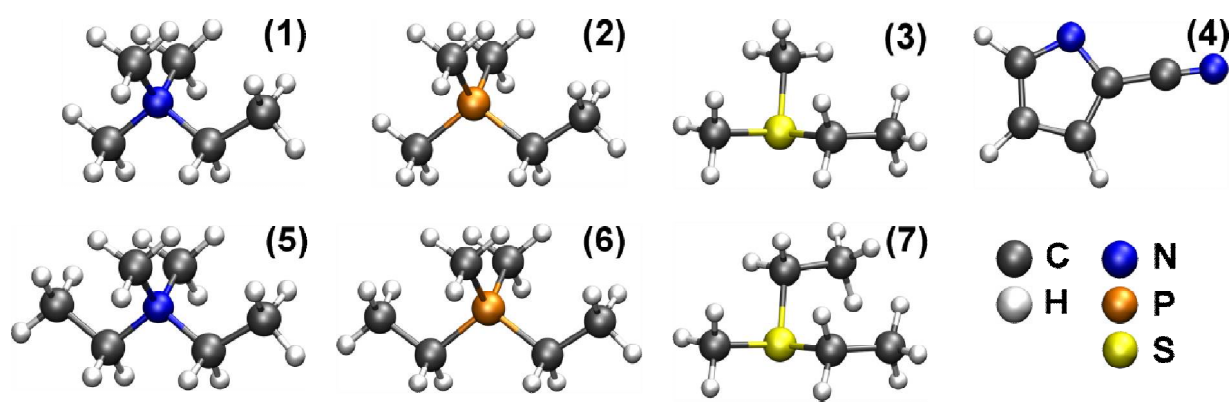

Figure 1. Optimized geometries of one AHA and six weakly coordinating cations used to construct RTILs in the present work: (1) ethyltrimethylammonium $[N_{2111}]^+$; (2) ethyltrimethylphosphonium $[P_{2111}]^+$; (3) ethyldimethylsulfonium $[S_{211}]^+$; (4) 2-cyanopyrrolidine [2-CNpyr]$^-$; (5) diethyldimethylammonium $[N_{2211}]^+$; (6) diethyldimethylphosphonium $[P_{2211}]^+$; (7) diethylmethylsulfonium $[S_{221}]^+$.

Significant progress was achieved in capturing $CO_2$ by the functionalized RTILs within recent years.[15-28] Chen and coworkers[15] studied (3-aminopropyl)tributylphosphonium glycinate in the aqueous medium. These researchers found that $CO_2$ uptake occurs due to the neutralization reaction of the dual amino bases. In turn, Kim and coworkers[16] reported equimolar $CO_2$ absorption capacities of the primary and secondary amino-functionalized RTILs. Although amino-functionalized RTILs usually improve $CO_2$ absorption capacity,[29,30] they also considerably increase the viscosity of the corresponding RTIL.[17,18]

A number of non-amino RTILs based on the quaternary phosphonium cations were successfully tested for $CO_2$ capture.[19-22] Thirteen RTILs, containing aprotic heterocyclic anions paired with tetraalkylphosphonium- and imidazolium-based cations were studied by Fillion and coworkers.[22] Atilhan and coworkers[19] demonstrated that the anion plays a paramount role in determining $CO_2$ solubility using tributylmethylphosphonium formate, butyltrimethylammonium bis(trifluoromethanesulfonyl)imide, 1-methyl-1-propylpyrrolidinium dicyanamide, and 1-ethyl-3-methylimidazolium acetate. Jacquemin and coworkers[23] found that the solubilities of $CO_2$ in RTILs with the butyltrimethylammonium cation are higher than in the imidazolium-based RTILs. The effect of the alkyl chain length in the tetraalkylphosphonium cation has been recently debated by several groups.[20,22,24,25] The entropy-associated factors were found to enhance the $CO_2$ binding affinity in the phosphonium-based RTILs with shorter hydrophobic chains.[25]

The advantages of pairing the phosphonium-based cations with AHA were ascribed to the chemisorption mechanism.[25-27,31] The chemical interactions of the azolide anion, studied previously by Gurkan and coworkers,[28] were complemented by the $CO_2$ reaction with the cation (Figure 2). The mechanism encompassing an ylide intermediate[27] was proposed. The proton leaves α-carbons of the cations and gets substituted by the carboxyl group. Ultimately, the proton joins the AHA.

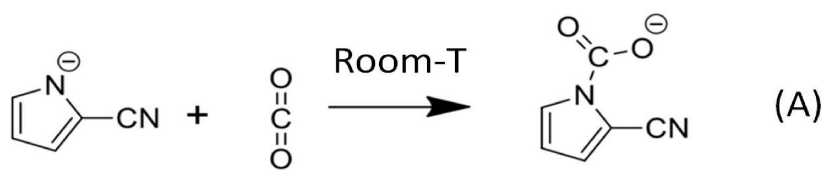

(A)

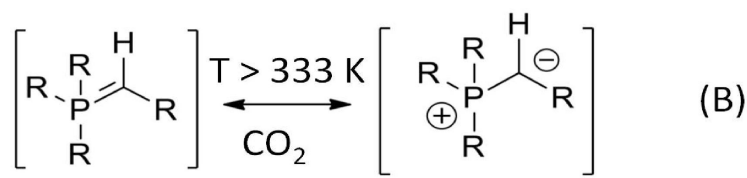

(B)

Figure 2. (A) The room-temperature chemical sorption of $CO_2$ by the anion. (B) The elevated-temperature ylide formation induced by $CO_2$.

The carbon dioxide absorption reaction kinetics[26] and enthalpy[31] for the trihexyltetradecylphosphonium cation-based RTILs are available. It was shown that the reaction enthalpy varies from -37 to -54 kJ mol$^{-1}$ depending on the AHA structure. The carboxylation of the hydrophobic chain of the cation is kinetically limited at ambient temperature. However, it occurs at elevated temperatures in some RTILs.[27,31] The $CO_2$ sorption by the tetrabutylphosphonium 2-cyanopyrrolidine ionic liquid reaches 60 mole/mole % at 60$^o$C and 1 atm.[31] In addition to the previously experimentally reported phosphonium-based cations, we presently investigate ammonium- and sulfonium-based weakly coordinating cations. Furthermore, we consider the feasibility of the additional chemisorption reactions, such as simultaneous carboxylation of two α-carbon atoms and carboxylation of the β-carbon atoms of the ammonium-, phosphonium, and sulfonium-based cations. The results including thermodynamic potentials, structure and electronic properties supplemented by the infra-red spectra are fundamentally insightful. They facilitate an interpretation of the experimental data and outline principal possibilities for the extension of the carbon dioxide scavenging procedures.

This work reports a few $CO_2$ chemisorption reactions for the following RTILs: ethyltrimethylammonium 2-cyanopyrrolidine [N$_{2111}$][2-CNpyr], ethyltrimethylphosphonium 2-cyanopyrrolidine [P$_{2111}$][2-CNpyr], ethyldimethylsulfonium 2-cyanopyrrolidine [S$_{211}$][2-CNpyr], diethyldimethylammonium 2-cyanopyrrolidine [N$_{2211}$][2-CNpyr], diethyldimethylphosphonium 2-cyanopyrrolidine [P$_{2211}$][2-CNpyr], and diethylmethylsulfonium 2-cyanopyrrolidine [S$_{221}$][2-CNpyr]. In addition to the experimentally exemplified carboxylation mechanism that involves a ylide intermediate[32] and attaches $CO_2$ to the α-carbon atom of each cation, we consider the carboxylation of the β-carbon atom and the carboxylation of the two α-carbon atoms of each cation. The obtained thermochemical information can be readily used to interpret the existing $CO_2$ capture experiments and schedule new trials at possibly more aggressive conditions.

**Computational Systems and Details**

The enthalpies and Gibbs free energies at a set of temperatures and pressures were derived using the composite thermochemistry method Gaussian-4 proposed by Curtiss and coworkers[33] that includes eight steps of subsequent wave function-based calculations at different levels of theory. The list of the simulated systems is given in Table 1.

First, the local-minimum geometries of the simulated systems were obtained at the hybrid density functional level. Second, accurate vibrational frequencies were computed. The additional stages of the thermochemical calculation are needed to introduce a range of corrections, for instance, to account for the correlations of electrons. Very accurate potential energies of the systems are critical to ultimately obtain reliable thermodynamic potentials. Based on previous benchmarking provided by Curtiss and coworkers,[33] an average discrepancy between the calculated thermochemical data and the corresponding experimental data does not exceed 1 kcal mol$^{-1}$.

Table 1. Abbreviations, empirical formulae, and brief descriptions of the simulated RTILs probed for carbon dioxide capture.

| Cation | Notation | Chemical Formula | Comments |
|---|---|---|---|
| | N0 | $[NC_5H_{14}][N_2C_5H_3]$ | Reactant $[N_{2111}][2\text{-CNpyr}]$ |
| Ammonium-based | N1 | $[NC_6H_{16}][N_2C_5H_3]$ | Reactant $[N_{2211}][2\text{-CNpyr}]$ |
| | N2 | $[NC_6H_{13}O_2][N_2C_5H_4]$ | $CO_2$ attached to α-carbon |
| | N3 | $[NC_6H_{13}O_2][N_2C_5H_4]$ | $CO_2$ attached to β-carbon |

| | N4 | $[NC_8H_{14}O_4][N_2C_5H_4]$ | $CO_2$ attached to 2 α-carbons |
|---|---|---|---|
| Phosphonium-based | P0 | $[PC_5H_{14}][N_2C_5H_3]$ | Reactant $[P_{2111}][2\text{-CNpyr}]$ |
| | P1 | $[PC_6H_{16}][N_2C_5H_3]$ | Reactant $[P_{2211}][2\text{-CNpyr}]$ |
| | P2 | $[PC_6H_{13}O_2][N_2C_5H_4]$ | $CO_2$ attached to α-carbon |
| | P3 | $[PC_6H_{13}O_2][N_2C_5H_4]$ | $CO_2$ attached to β-carbon |
| | P4 | $[PC_8H_{14}O_4][N_2C_5H_4]$ | $CO_2$ attached to 2 α-carbons |
| Sulfonium-based | S0 | $[SC_4H_{11}][N_2C_5H_3]$ | Reactant $[S_{211}][2\text{-CNpyr}]$ |
| | S1 | $[SC_5H_{13}][N_2C_5H_3]$ | Reactant $[S_{221}][2\text{-CNpyr}]$ |
| | S2 | $[SC_5H_{10}O_2][N_2C_5H_4]$ | $CO_2$ attached to α-carbon |
| | S3 | $[SC_5H_{10}O_2][N_2C_5H_4]$ | $CO_2$ attached to β-carbon |
| | S4 | $[SC_6H_{11}O_4][N_2C_5H_4]$ | $CO_2$ attached to 2 α-carbons |

Hess' law rigorously states that enthalpy changes are always additive. Therefore, the reaction enthalpy does not depend on the reaction path. The concepts of the Hess' law can be expanded to include changes in entropy and in Gibbs free energy, as they are also the functions of state. Based on this theory, the respective thermodynamic potentials for the chemisorption reactions can be derived as a difference of the products' potential of state and the reactants' potential of state.

For the calculation of all other reported properties, the following computational procedures apply. The B3LYP hybrid density functional[34,35] was used to optimize the wave function. The basis functions of the atom-centered polarized triple-zeta split-valence set with diffuse functions for all atoms 6-311++G** were employed. The wave function convergence criterion was set to $10^{-8}$ Hartree. The geometry optimization was carried out according to the Berny algorithm. The four geometry convergence criteria of 12 kJ nm$^{-1}$ for the maximum force, 8 kJ nm$^{-1}$ for the root-mean-squared force, 0.18×10$^{-3}$ nm for the maximum displacement, and 0.12×10$^{-3}$ nm the root-mean-squared displacement were simultaneously used. The partial charges of the atoms were obtained according to the definition of Hirshfeld.[36] In turn, the profiles of vibrational frequencies obtained in the thermochemistry calculations were used to derive infrared spectra.

The described above electronic-structure calculations were conducted in Gaussian 09.[37] The optimized geometries were visualized, analyzed, and the corresponding molecular images were prepared in the VMD-1.9 software.[38]

**Results and Discussion**

Figure 3 provides the schemes of the chemisorption reactions in which a single $CO_2$ molecule transforms into the carboxyl group linked to the α-carbon atom. Next, the carboxyl groups donors its proton to [2-CNpyr]⁻, i.e. the AHA anion. Note that the products of such chemisorption reactions are two on the whole neutral particles. The transformation of ions into molecules is an important feature of these reactions clearly contributing to their thermodynamic allowance since any molecules are always more stable than their conjugated cations and anions. The acyclic cations give upon chemisorption rise to rather unusual zwitterionic structures.

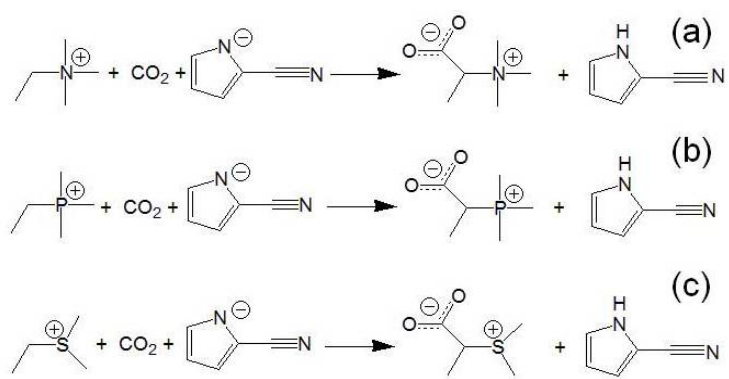

Figure 3. The schemes summarizing $CO_2$ grafting to α-carbon atoms of the investigated acyclic cations: (a) $[N_{2111}]^+$; (b) $[P_{2111}]^+$; (c) $[S_{211}]^+$. The studied transformations of the RTILs (Table 1) are as follows: N0→N2, P0→P2, S0→S2.

In this paper, we concentrate our efforts on the carboxylation of the acyclic weakly coordinating cations $[N_{2111}]^+$, $[P_{2111}]^+$, $[S_{211}]^+$, $[N_{2211}]^+$, $[P_{2211}]^+$, $[S_{221}]^+$ since their role as $CO_2$ scavengers is rather uncertain. In turn, the successful performance of AHA occurs through the carbamate formation mechanism, hence, we do not compute the thermodynamic potentials for the corresponding subprocess. It is essential to note that the specific structure of the proton acceptor does not influence the performance or the considered acyclic cations. For instance, instead of $[2\text{-CNpyr}]^-$, halogenide ionic liquids, such as tetrabutylammonium chloride, tetrabutylphosphonium bromide, tetrabutylammonium iodide and others, can readily be used.

The carboxylation of the cations takes place through the ylide intermediate formation in the presence of $CO_2$ and elevated temperature (Figure 2). This mechanism fosters the chemical reaction which would be impossible due to the energetic barrier at other conditions. Indeed, the computed thermochemistry for the deprotonation of the alkyl chain is strongly positive, +1134 kJ mol$^{-1}$ for $[N_{2111}]^+$, +1116 kJ mol$^{-1}$ for $[P_{2111}]^+$ and +1108 kJ mol$^{-1}$ for $[S_{211}]^+$. Note that deprotonation of $[N_{2111}]^+$ is the least thermodynamically favorable process. All values are of the same order meaning that carboxylation in the case of the first sorbed $CO_2$ molecule occurs similarly. In turn, the protonation of the 2-cyanopyrrolidine anion contributes -1412 kJ mol$^{-1}$ to the overall reaction. On the aggregate, the $CO_2$ capture is thermodynamically favorable. Recall that the ylide formation greatly decreases the reaction barrier.[33]

Figure 4 provides Gibbs free energies for the N0→N2, P0→P2, S0→S2 reactions (α-carbon target) and the N0→N3, P0→P3, S0→S3 reactions (β-carbon target) at ambient conditions, whereas Figure 5 attempts to correlate them to electrophilicities of the corresponding carbon atoms.

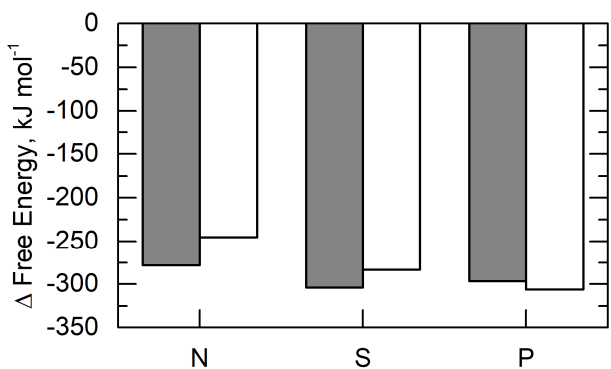

Figure 4. The standard Gibbs free energies for the carboxylation chemisorption reactions at α-carbon atom (N0→N2, P0→P2, S0→S2, gray bars) and β-carbon atom (N0→N3, P0→P3, S0→S3, white bars) of the cations. The x-axis is a central atom of the aprotic acyclic cation.

When $CO_2$ attaches to the α-carbon atom of the sulphonium-based cation, the Gibbs free energy of the overall reaction is most favorable. However, the deviation from the phosphonium- and ammonium-based cations is not drastic. In the case of the β-carbon targets, the phosphonium cation provides the most favorable reaction energetics. At both targets, the ammonium-based cation performs worse than its competitors. All chemisorption reactions at the acyclic weakly coordinating cations are moderately thermodynamically allowed.

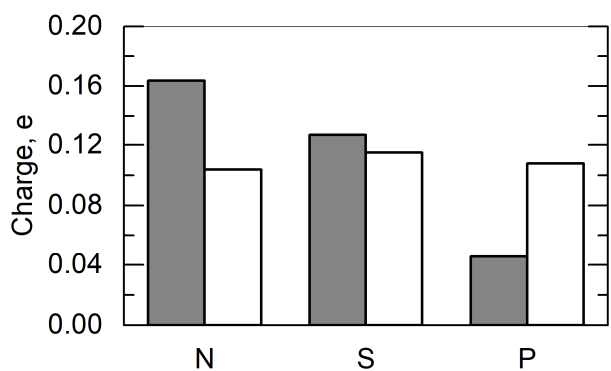

Figure 5. The distribution of the partial charges over the alkyl chain's carbon atoms depending on the central atom of the cation: gray bars correspond to α-carbon atoms, white bars correspond to β-carbon atoms. The partial charges were computed according to the Hirshfeld algorithm[36] by ascribing the electron density of the hydrogen atoms to the closest covalently linked heavy atoms. The x-axis is a central atom of the aprotic acyclic cation.

By analyzing the partial point charges of the reactive sites of the cations (Figure 5), we found that it is impossible to explain the effect of the cationic reactant by the hydrophilicity of the respective grafting site following the earlier elaborated methodology.[29,30] Compare the electronegativity of nitrogen (3.04) to the electronegativity of sulfur (2.58) to the electronegativity of phosphorus (2.19).[39] A more electropositive cationic center donates a larger fraction of the electron density to the α-carbon atoms. This is clearly seen in Figure 5. The α-carbon atom of $[P_{2111}]^+$ carries a smaller positive charge than the α-carbon atoms of $[S_{211}]^+$ and $[N_{2111}]^+$. In turn, the β-carbon atoms are nearly insensitive to the electronegativity of the cation's central atom, compare the partial Hirshfeld charges[36] in Figure 5(b) to one another.

While it may be expected that $CO_2$, a nucleophilic particle, prefers more electron-deficient sites of the acyclic cations, the computed electrochemistry for the aggregated chemisorption reactions does not confirm this guess. The ammonium-based cation performs more poorly than other cations although it supplies the most electrophilic α-carbon atom. The dimethylethylsulfonium cation performs 8 kJ mol$^{-1}$ better than the ethyltrimethylphosphonium cation at the α-carbon site. On the contrary, the dimethylethylsulfonium cation appears 23 kJ mol$^{-1}$ worse than the ethyltrimethylphosphonium cation in the case of the β-carbon atom carboxylation because the α-carbon atom of $[S_{211}]^+$ is somewhat more nucleophilic. All in all, the carboxylation of the nor α-, neither β-carbon atoms can be described by the factor of nucleophilicity. Indeed, the larger positive partial charge does not correspond to a more favorable Gibbs free energy.

The above unusual observations lead us to the hypothesis that steric factors are likely involved in the carboxylation of the cationic α-carbon and possibly β-carbon atoms. Figure 6 visualizes the most representative distances and partial electronic charges in the S2 and S3 structures (see Table 1). Note the difference in the sulfur-oxygen non-covalent distances, 0.23 nm vs. 0.26 nm. These atoms possess quite large charges resulting in a significant fraction of the intramolecular electrostatic attraction. Such an attraction contributes directly to enthalpy and Gibbs free energy of the corresponding reactions. Apart from the electrostatic attraction, the electron-electron repulsion comes into play, especially in the case of the β-carbon reaction site. Indeed, the equilibrium sulfur-oxygen non-bonded distance is convincingly smaller than the sum of the van der Waals radii of the involved chemical elements.[39]

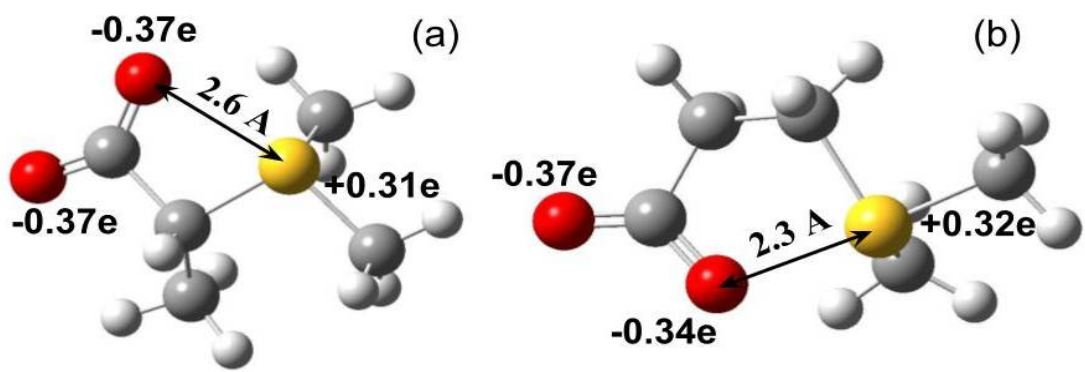

Figure 6. The partial charges and the selected geometrical parameters for the chemisorption products: (a) S2 and (b) S3. Oxygen atoms are red; carbon atoms are grey; hydrogen atoms are white; the sulfur atom is yellow.

An interplay of the enumerated factors results in similar Gibbs free energies for both carboxylation sites in the sulfonium-based cation $[S_{211}]^+$. Similar structural and partial charge distribution patterns were also observed in the case of the ammonium- and phosphonium-based cations. In $[N_{2111}]^+$, the nitrogen charge is +0.05 a.u. and the nitrogen-oxygen distance is 0.28 nm. The carboxyl oxygen's charge amounts to -0.37 a.u. These enumerated parameters correspond to the α-carbon reaction site. When chemisorption takes place at the β-carbon site, the partial charge on nitrogen is +0.06 a.u., whereas the charge of oxygen is -0.40 a.u. The nitrogen-oxygen distance is 0.33 nm. From this data, one can conclude that a small partial electronic charge on the nitrogen atom is a major reason for the relatively poor chemisorption performance of the ethyltrimethylammonium cation. Furthermore, it is interesting to compare the carboxyl oxygen–central atom of the cation non-covalent distances which are significantly different upon the carboxylation of the investigated cations. The thermodynamic stabilization of the products via the oxygen-nitrogen, oxygen-sulfur, and oxygen-phosphorus electrostatic interactions looks very important to rationalize the different performances of the ternary and quaternary cations.

The case of the phosphonium-based cation supports the above-argued model. The phosphorus' charge is +0.47 a.u., whereas the oxygen's charge is -0.37 a.u. The resulting electrostatic attraction forms the phosphorus-oxygen non-covalent distance of 0.26 nm. This data corresponds to the α-carbon carboxylation site. In turn, when the β-carbon reaction site is considered, the partial charges are +0.45 a.u. (phosphorus) and -0.28 a.u. (oxygen), whereas the phosphorus-oxygen non-covalent distance is 0.21 nm. Such a small distance is a very convincing explanation of why $[P_{2111}]^+$ performs better than $[S_{211}]^+$ for the case of the β-carbon reaction site. The geometry of the carboxyl group which is rigorously predefined by the hybridization state of its central atom leads to the sterical hindrance and induces an electron-electron electrostatic repulsion. While the covalent radii of phosphorus and sulfur are very similar, the smaller number of the alkyl chains (Figure 1) in $[S_{211}]^+$ and $[S_{221}]^+$ as compared to $[P_{2111}]^+$ and $[P_{2211}]^+$ allow more free space for the $CO_2$ molecule to accommodate conformational changes and transform into the carboxyl group in the case of α-carbon site. The accessibility of β-carbon in less affected by the number of alkyl chains, and the high phosphorous atom charge becomes decisive.

The strong intramolecular electrostatic interactions represent a rather exotic observation in organic chemistry. Their emergence at notably small distances, 0.21-0.33 nm, leads to the perturbations of the ionic and molecular structures. Due to the distortions of the functional groups, the chemical particle undergoes significant alterations in its structure. We analyzed the changes of the α-carbon-carboxyl carbon single-covalent bond lengths in the chemisorption products. In the case of the α-carbon carboxylation site, the bond lengths are as follows: 0.160 nm in $[N_{2111}]^+$, 0.159 nm in $[P_{2111}]^+$, 0.159 nm in $[S_{211}]^+$. In the case of the β-carbon carboxylation site, the bond lengths are systematically shorter: 0.157 nm in $[N_{2111}]^+$, 0.154 nm in $[P_{2111}]^+$, 0.156 nm in $[S_{211}]^+$. Recall that the conventional single bond carbon-carbon length is 0.152 nm. The bond elongation which was observed in all carboxylation products should be ascribed to the intramolecular electrostatic interactions and sterical hindrances.

Figure 7 provides the standard enthalpies of the chemisorption reactions, the corresponding schemes are shown in Figure 3. All chemisorption processes are moderately exothermic and in line with the Gibbs free energies discussed above. The least exothermic process occurs in the case of the ammonium-based cation $[N_{2111}]^+$. Its thermochemistry does not depend on the identity of the alkyl chain carbon atom. Whereas $[S_{211}]^+$ performs somewhat better than $[P_{2111}]^+$ for grafting at the α-carbon atom, the trend reverses for the grafting at the β-carbon atom. The identified energy differences are, however, relatively small. They can be interpreted in exactly the same way as we have already exemplified with the Gibbs free energies.

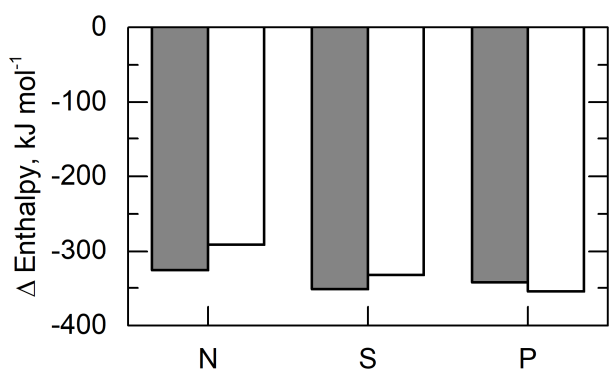

Figure 7. The standard enthalpies corresponding to the $CO_2$ chemisorption reactions at the α-carbon atoms (N0→N2, P0→P2, S0→S2, gray bars) and β-carbons atoms (N0→N3, P0→P3, S0→S3, white bars) of the cations. The x-axis is a central atom of the aprotic acyclic cation.

Comparison of the computed standard enthalpies and Gibbs free energies reveals that all chemisorption reactions are strongly enthalpically driven. Nevertheless, the role of the entropic factor is clear based on general chemical wisdom. Attachment of $CO_2$ to the alkyl chain of the cation is accompanied by significant conformational alterations. First, the $-CO_2$ moiety loses the linear geometry. Second, this group of atoms acquires a negative electrostatic charge. For these transformations to take place, certain free space is required.

Figure 8 depicts a scheme in which two $CO_2$ molecules simultaneously attach to the two α-carbon atoms of the aprotic acyclic cations, i.e. two alkyl chains of the cations acquire the acidic groups. In the following discussion, we will refer to this chemisorption mechanism as double carboxylation as opposed to the single carboxylation that we characterized earlier. Since the carboxylation of the cationic alkyl chain is followed by the deprotonation to donate the proton to AHA, the cation becomes a molecule upon the first stage and an anion upon the second stage. The AHA is, in the meantime, transformed into a molecule.

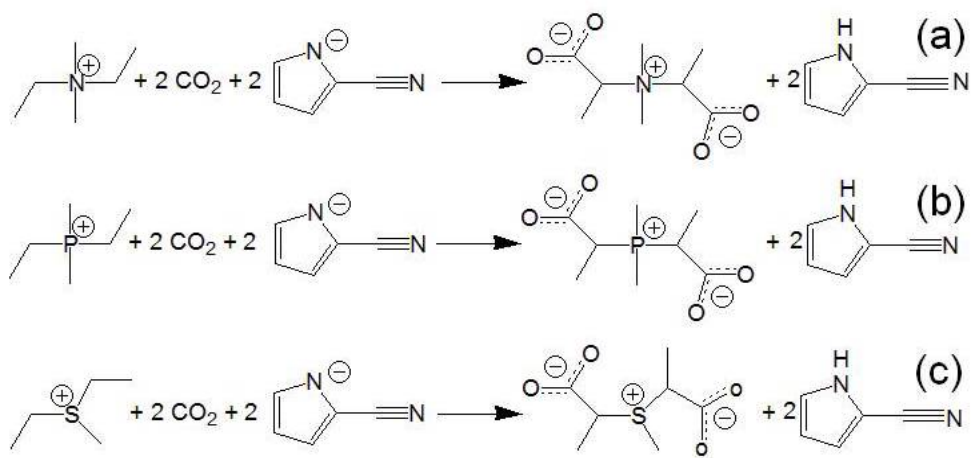

Figure 8. The schemes summarizing $CO_2$ grafting to two α-carbons of the investigated acyclic cations simultaneously: (a) diethyldimethylammonium $[N_{2211}]^+$; (b) diethyldimethylphosphonium $[P_{2211}]^+$; (c) diethylmethylsulfonium $[S_{221}]^+$. The studied transformations of the RTILs (Table 1) are as follows: N1→N4, P1→P4, S1→S4.

Interestingly, the double carboxylation process is thermodynamically allowed (Figure 9) that can be scarcely predicted without the specific computational investigation. First, there exists a minimum point (no imaginary frequencies in the vibrational profile) on the potential energy surface corresponding to the product of double carboxylation. Second, the enthalpies and Gibbs free energies of all chemisorption reactions are moderately negative indicating spontaneous processes. We have no reasons to doubt that the kinetic mechanism of double carboxylation inherits the ylide intermediate pathway that was identified in the synthetic work.[32]

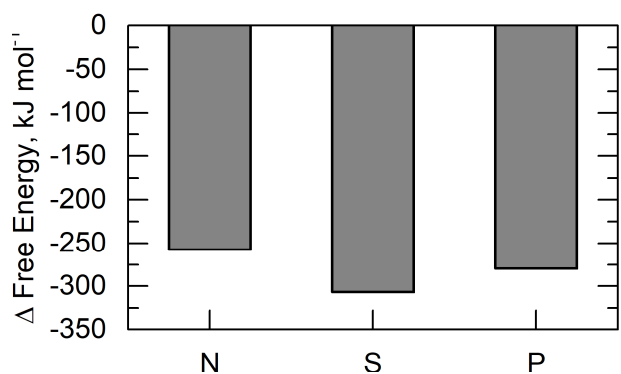

Figure 9. The standard Gibbs free energies for the carboxylation chemisorption reactions at two α-carbon atoms simultaneously (N1→N4, P1→P4, S1→S4) of the cations. The x-axis is a central atom of the aprotic acyclic cation.

The geometry and selected partial Hirshfeld charges[36] of the products (see N4, P4, S4 in Table 1) obtained from the $[N_{2211}]^+$ and $[P_{2211}]^+$ cations are shown in Figure 10. In system N4, the charge on the nitrogen atom is +0.05 a.u., whereas the carboxyl oxygen's charges are 0.40-0.41 a.u. The phosphorus atom carries the charge of +0.44 a.u. in system P4. For the sake of understanding the thermochemistry trends, it is interesting to compare the latter to the partial charge in system S4, being +0.28 a.u. The oxygen-nitrogen and oxygen-phosphorus non-covalent distances amount to 0.29 nm. The oxygen-sulfur distance derived from system S4 is 0.28 nm. These distances indicate strong electrostatic attraction interactions that stabilize P4 and S4 to a larger extent and N4 to a lesser extent in accordance with their partial charges in the chemisorption products.

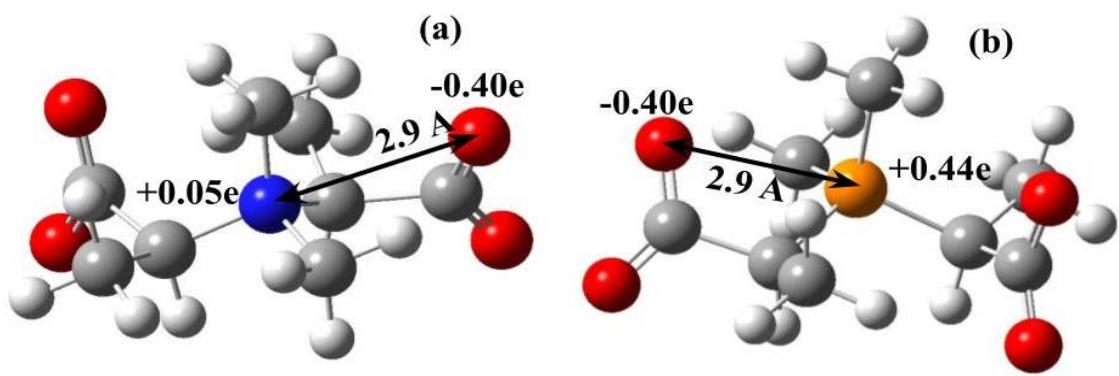

Figure 10. The partial charges and geometrical parameters for the chemisorption products: (a) N4 and (b) P4. Oxygen atoms are red; carbon atoms are grey; hydrogen atoms are white; the nitrogen atom is blue; the phosphorus atom is orange.

The computed standard Gibbs free energies of the double carboxylation are in line with the previously discussed data and can be rationalized by simultaneously considering sterical hindrances, electrostatic intrazwitterionic attractions, and nucleophilicies of the carbon atoms of the cations.

In addition to featuring the smallest oxygen-central atom of the cation distance of 0.28 nm, system S4 induces fewer distortions to the carboxyl groups. Indeed, the sulfonium-based cation contains only three alkyl chains, whereas $[P_{2111}]^+$ and $[N_{2211}]^+$ contain four chains each. In the meantime, the atomic radii of phosphorus and sulfur differ insignificantly.[39]

Table 2 represents the dependences of the Gibbs free energies on temperature and pressure for the reactions at the α-carbon site, β-carbon sites, and double carboxylation. The increase of pressure is favorable for all chemisorption reactions since a higher pressure value corresponds to a smaller entropic contribution in the simulated systems. The effect of temperature is opposite because the reactions are exothermic (Figure 4). In the case of the α-carbon reaction sites, the performance of the cations is as follows: $[N_{2111}]^+ < [P_{2111}]^+ < [S_{211}]^+$, whereas $[P_{2111}]^+$ performs systematically better than $[N_{2111}]^+$. In the case of the β-carbon reaction sites, the performance of the cations is as follows: $[N_{2111}]^+ < [S_{211}]^+ < [P_{2111}]^+$. In the case of the double carboxylation, the performance of the cations is as follows: $[N_{2211}]^+ < [P_{2211}]^+ < [S_{221}]^+$.

Table 2. The Gibbs free energies corresponding to the single and double chemisorption reactions at non-standard conditions: (1) temperatures of 350 K, 400 K; 450 K and (2) pressures of 1 bar, 50 bar, 100 bar, 150 bar, 200 bar for all investigated RTILs. All thermodynamic potentials are given per mole of carbon dioxide.

|  | Pressure, bar | | | | |
|---|---|---|---|---|---|
| T, K | 1.00 | 50.0 | 100 | 150 | 200 |
| Gibbs free energy, kJ mol$^{-1}$ | | | | | |
| CO$_2$ grafting to the α-carbon atom of the $[N_{2111}]^+$ cation | | | | | |
| 350 | -270 | -281 | -283 | -284 | -285 |
| 400 | -262 | -275 | -277 | -278 | -279 |
| 450 | -254 | -268 | -271 | -272 | -273 |
| CO$_2$ grafting to the β-carbon atom of the $[N_{2111}]^+$ cation | | | | | |
| 350 | -238 | -249 | -251 | -252 | -253 |
| 400 | -230 | -243 | -245 | -247 | -248 |
| 450 | -222 | -237 | -240 | -241 | -242 |
| CO$_2$ grafting to the two α-carbon atoms of the $[N_{2211}]^+$ cation | | | | | |
| 350 | -121 | -132 | -134 | -135 | -136 |
| 400 | -112 | -125 | -128 | -129 | -130 |
| 450 | -104 | -119 | -122 | -123 | -124 |
| CO$_2$ grafting to the α-carbon atom of the [S$_{211}$]$^+$ cation | | | | | |
| 350 | -296 | -307 | -309 | -310 | -311 |
| 400 | -288 | -301 | -303 | -304 | -305 |
| 450 | -280 | -295 | -297 | -299 | -300 |
| CO$_2$ grafting to the β-carbon atom of the [S$_{211}$]$^+$ cation | | | | | |
| 350 | -274 | -286 | -288 | -289 | -290 |
| 400 | -266 | -279 | -281 | -283 | -284 |
| 450 | -258 | -273 | -275 | -277 | -278 |
| CO$_2$ grafting to the two α-carbon atoms of the [S$_{221}$]$^+$ cation | | | | | |
| 350 | -146 | -157 | -159 | -160 | -161 |
| 400 | -138 | -151 | -153 | -155 | -156 |
| 450 | -131 | -148 | -148 | -149 | -150 |
| CO$_2$ grafting to the α-carbon atom of the [P$_{2111}$]$^+$ cation | | | | | |
| 350 | -288 | -300 | -302 | -303 | -304 |
| 400 | -281 | -294 | -296 | -298 | -298 |
| 450 | -273 | -288 | -290 | -292 | -293 |
| CO$_2$ grafting to the β-carbon atom of the [P$_{2111}$]$^+$ cation | | | | | |
| 350 | -298 | -309 | -311 | -312 | -313 |
| 400 | -290 | -303 | -305 | -306 | -307 |
| 450 | -282 | -296 | -299 | -300 | -301 |
| CO$_2$ grafting to the two α-carbon atoms of the [P$_{2211}$]$^+$ cation | | | | | |
| 350 | -132 | -143 | -145 | -146 | -147 |
| 400 | -124 | -137 | -139 | -141 | -142 |
| 450 | -116 | -131 | -133 | -135 | -136 |

Mid- and far-infrared spectra for the pristine and carboxylated cations are provided in Figures 11-13. The carboxylation of the alkyl chain of the cations produces a strong carbon-oxygen stretching band. Furthermore, it introduces significant perturbations to the structure of former the cation resulting in a shift of numerous peaks. The carbon-hydrogen stretching vibration at over 3000 cm$^{-1}$ in the modified methylene group of each cation is a convincing fingerprint of the successful carboxylation reaction.

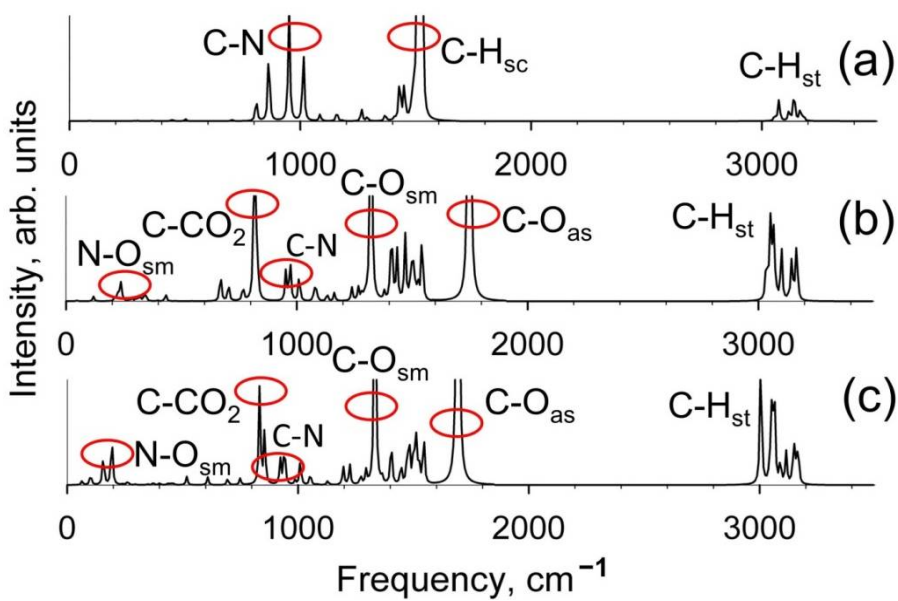

Figure 11. Mid- and far-infrared vibrational spectra of (a) N0 (non-modified cation); (b) N2; and (c) N3 systems. Abbreviations: "sm" – symmetric stretching; "as" – anti-symmetric stretching; "sc" – scissoring; "st" – stretching.

The carbon-nitrogen bond of $[N_{2111}]^+$, system N0, stretches at 950 cm$^{-1}$ (Figure 11). After $CO_2$ is grafted, a substantial redshift is seen, towards 828 cm$^{-1}$. The newly formed covalent bond between the α-carbon of $[N_{2111}]^+$ and the carbon atom of the carboxyl group exhibits symmetric stretching vibration at 816 cm$^{-1}$. This frequency can be used to identify such bonds in laboratory experiments. The frequencies of 1322 cm$^{-1}$ and 1747 cm$^{-1}$ in system N2 correspond to -COO$^-$. Remarkably, the non-covalent oxygen (COO$^-$)–nitrogen of $[N_{2111}]^+$ electrostatic interaction engenders symmetric stretching vibration at 239 cm$^{-1}$.

After $CO_2$ joins the cation at the β-carbon position, system N3, the carbon-nitrogen band shifts to 855 cm$^{-1}$. The symmetric stretching of the α-carbon–carboxylic carbon bond shifts to 834 cm$^{-1}$. Carbon-oxygen vibrations shift to 1333 cm$^{-1}$ (symmetric stretching) and 1692 cm$^{-1}$ (asymmetric stretching). The previously mentioned vibration due to the strong Coulombic attraction was observed at 193 cm$^{-1}$.

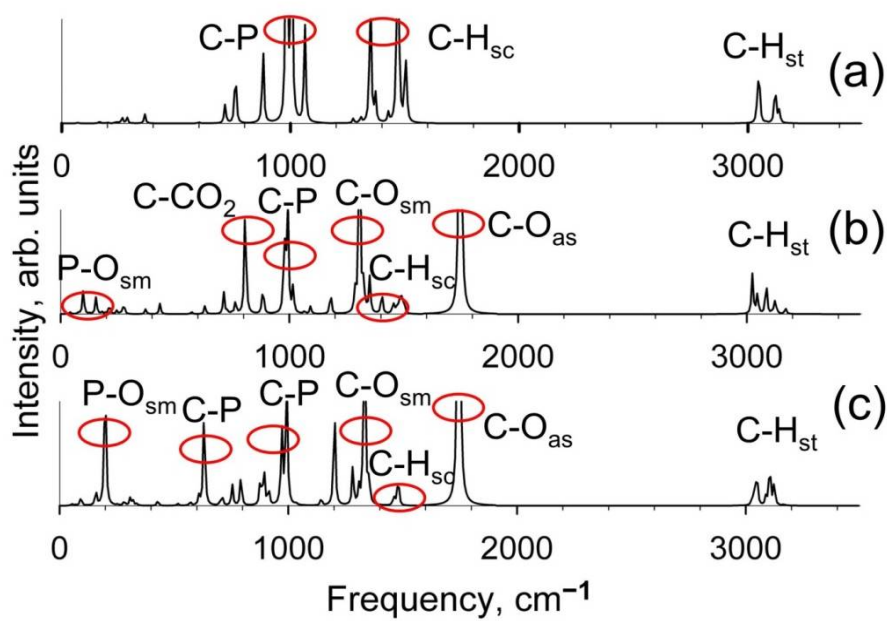

Figure 12. Mid- and far-infrared vibrational spectra of the (a) P0 (non-modified cation), (b) P2, and (c) P3 systems. Abbreviations: "sm" – symmetric stretching; "as" – anti-symmetric stretching; "sc" – scissoring; "st" – stretching.

Figure 12 reports the spectrum for the pristine ethyltrimethylphosphonium cation and its derivatives after the $CO_2$ capture, systems P0, P2, and P3. The polar covalent bond that connects the nitrogen and α-carbon atoms of the cation vibrates at 1005 cm$^{-1}$. If $CO_2$ is grafted to the α-carbon atom, system P2, the frequency is somewhat modified, 992 nm$^{-1}$. The symmetric stretching band of the α-carbon – carboxylic carbon bond is seen at 807 cm$^{-1}$. Similar to the case of ethyltrimethylammonium, the oxygen-phosphorus symmetric stretching was observed at 99 cm$^{-1}$. Upon $CO_2$ grafting to the β-carbon atom, system P3, the carbon-phosphorus bond frequency becomes 632 cm$^{-1}$. In turn, the α-carbon-carboxyl carbon bond stretches symmetrically at 894 cm$^{-1}$. The non-covalent vibration that involves carboxyl oxygen and phosphorus was found at 200 cm$^{-1}$.

Figure 13 provides the vibrational spectrum of systems S0, S2, and S3. In the pristine sulfonium-based cation, the sulfur-α-carbon bond vibrates at 662 cm$^{-1}$. This frequency decreases to 605 cm$^{-1}$ in the carboxylated state, system S1. The carbon-carboxyl carbon bond signals through symmetric stretching at 804 cm$^{-1}$. The carboxyl group vibrates symmetrically at 1311 cm$^{-1}$ and asymmetrically at 1752 cm$^{-1}$, systems S2 and S3. Ultimately, the characteristic non-covalent carboxyl-sulfur symmetric stretching that essentially influences thermochemistry was registered at 186 cm$^{-1}$.

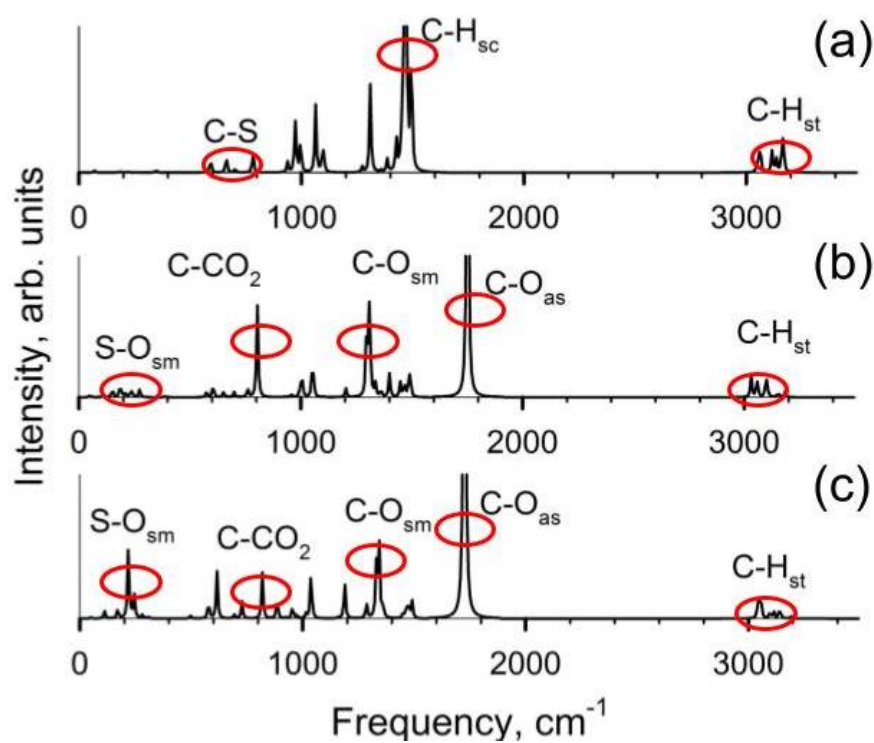

Figure 13. Mid- and far-infrared vibrational spectra of the (a) S0 (non-modified cation), (b) S2 and (c) S3 systems. Abbreviations: "sm" – symmetric stretching; "as" – anti-symmetric stretching; "sc" – scissoring; "st" – stretching.

**Conclusions**

We hereby performed a comprehensive theoretical investigation of the principal possibility to use acyclic aprotic organic cations coupled with the 2-cyanopyrrolidine anion for the carbon dioxide capture and compared three different reaction sites that can be carboxylated through the ylide formation mechanism and further protonation of the anion. We computed thermochemistry, bonded and non-bonded geometrical parameters, charge density distribution in terms of partial atomic charges, and vibrational frequencies that can be experimentally recorded in the infrared spectroscopic range. The impact of the cationic nature was rationalized through nucleophilicity of the reaction sites and sterical hindrances that emerge after the chemisorption reaction takes place.

Due to the strongly favorable protonation of 2-cyanopyrrolidine, all studied carboxylation reactions are thermodynamically allowed. The chemisorption of carbon dioxide grafting to the aprotic acyclic cations follows the same mechanism. Still, it is possible to detect differences among reactants and link them to the nature of the central atoms and sterical hindrances to rationalize the reported thermochemical potentials. The electrostatic intramolecular attraction between the positively charged center (N, P, S) and the negatively charged carboxyl center plays an essential role in stabilizing the product and, therefore, fostering the chemisorption reaction. Since the opposite-signed charges are located very close to one another, 0.21-0.33 nm, the corresponding non-covalent interaction is of paramount importance.

The chemisorption products represent very unusual chemical structures of the zwitterionic nature. The attachment of the first carbon dioxide molecule transforms an acyclic aprotic cation into a neutral particle. The attachment of the second carbon dioxide molecule gives rise to a zwitterion with a negative total charge. Based on our simulations, these structures are relatively stable and may be isolated. The transformation of ions into molecules in the course of chemisorption is seen as a fruitful idea for chemical engineering.

The elevated pressure somewhat shifts the $CO_2$ capturing reaction equilibrium rightwards, whereas temperature increase is not desirable due to the exothermic behavior of the chemisorption process. The computed infrared spectra for modified and non-modified cations provide useful fingerprints that are characteristic of the carboxylated α-carbon and β-carbon atoms. The non-covalent electrostatic intramolecular attraction is clearly reflected by the vibrational profile at around 200 cm$^{-1}$.

The reported results, for the first time, rationalize the very high $CO_2$ capacities of the AHA containing phosphonium-based RTILs reported by Brennecke and coworkers.[27] We extend the experimental success onto the sulfonium-based RTILs and systematically rate the three families of RTILs based on their structural properties and electronic distribution profile. We argue that the phosphonium- and sulfonium-based RTILs exhibit a higher affinity to the carbon dioxide chemisorption as compared to the ammonium-based RTILs. According to the thermochemistry, the α-carbon and β-carbon atoms of the alkyl chains perform similarly in relation to $CO_2$ binding. The reported novel physical and chemical insights may be interesting to researchers who develop novel $CO_2$ environmentally benign gas scavengers. The provided rationalization and the corresponding methodology foster ongoing efforts to develop an alternative approach of $CO_2$ binding by the room-temperature ionic liquids through chemisorption.

**Acknowledgements**

The results of the work were obtained using computational resources of Peter the Great Saint-Petersburg Polytechnic University Supercomputing Center (www.spbstu.ru). I.V.V. would like to acknowledge the financial support by Project No. UID/QUI/50006/2021 (LAQV@REQUIMTE) with funding from FCT/MCTES through the Portuguese national funds.

**Contributions of the Authors**

V.V.C. formulated the final research schedule, provided mentorship for the team, interpreted the results, wrote the discussion and conclusions, and edited the entire manuscript. N.A.A. simulated the systems, prepared the figures and tables, and participated in the collaborative communication. I.V.V. treated the data, prepared the figures and tables, wrote an initial version of the introduction, and participated in the collaborative communication.

**Conflict of interest**

The authors hereby declare no financial interests that might bias the interpretations of the obtained results.

**Author for correspondence**

Inquiries regarding the scientific content of this paper shall be directed through electronic mail to Prof. Dr. Vitaly V. Chaban (vvchaban@gmail.com). Inquiries regarding the computational procedures should be directed to Dr. Nadezhda Andreeva (nadezhda.a.andreeva@gmail.com).